\documentclass[english,aps,floats,onecolumn,showpacs,nofootinbib]{revtex4}
\usepackage{pslatex}
\usepackage[T1]{fontenc}
\usepackage[latin1]{inputenc}
\usepackage{graphicx}
\usepackage{epsfig}

\usepackage{calc}
\usepackage{ifthen}

{
{
{
\newcommand{\bea}{\begin{eqnarray}}
\newcommand{\eea}{\end{eqnarray}}

\newcommand{\nc}{\newcommand}
\nc{\renc}{\renewcommand}
\nc{\eqs}[2]{\mbox{Eqs.~(\ref{#1},\,\ref{#2})}}
\nc{\eq}[1]{\mbox{Eq.~(\ref{#1})}}
\nc{\figs}[2]{\mbox{Figs.~(\ref{#1},\,\ref{#2})}}
\nc{\fig}[1]{\mbox{Fig~.(\ref{#1})}}
\nc{\be}[1]{\begin{equation} \mbox{$\label{#1}$}}
\nc{\ee}{\vspace{0.1cm}\end{equation}}

\newcommand{\bean}{\begin{eqnarray*}}
\newcommand{\eean}{\end{eqnarray*}}

\def\bfx{{\bf x}}
\def\bfk{{\bf k}}

\def\lae{\;^{<}_{\sim} \;} \def\gae{\; ^{>}_{\sim} \;}

\begin{document}
\title{Hemispherical Power Asymmetry from Scale-Dependent Modulated Reheating}
\author{John McDonald}
\email{j.mcdonald@lancaster.ac.uk}
\affiliation{Lancaster-Manchester-Sheffield Consortium for Fundamental Physics, Cosmology and Astroparticle Physics Group, Dept. of Physics, University of 
Lancaster, Lancaster LA1 4YB, UK}
\begin{abstract}

    We propose a new model for the hemispherical power asymmetry of the CMB based on modulated reheating. Non-Gaussianity from modulated reheating can be small enough to satisfy the bound from Planck if the dominant modulation of the inflaton decay rate is linear in the modulating field $\sigma$. $\sigma$ must then acquire a spatially-modulated power spectrum with a red scale-dependence. This can be achieved if the primordial perturbation of $\sigma$ is generated via tachyonic growth of a complex scalar field. Modulated reheating due to $\sigma$ then produces a spatially modulated and scale-dependent sub-dominant contribution to the adiabatic density perturbation.  We show that it is possible to account for the observed asymmetry while remaining consistent with bounds from quasar number counts, non-Gaussianity and the CMB temperature quadupole. The model predicts that the adiabatic perturbation spectral index and its running will be modified by the modulated reheating component.

\end{abstract}
\maketitle

\section{Introduction}

   The CMB temperature maps from WMAP \cite{wmapearly,wmap5} and Planck \cite{planckasym} show a hemispherical power asymmetry at the O(10)$\%$ level. The power asymmetry can be characterized by a temperature fluctuation dipole of the form \cite{gordon}
\be{x1} \frac{\delta T}{T}(\hat{n}) = s(\hat{n})\left[1 + A(\hat{n}.\hat{p}) \right]   ~,\ee
where $s(\hat{n})$ is a statistically isotropic map, $A$ is the magnitude of the asymmetry and $\hat{p}$ is its direction. The WMAP5 Internal Linear Combination (ILC) map found $A = 0.072 \pm 0.022$ in direction $(l, b) = (224, -22) \pm 24$ for multipoles $l \leq 64$ \cite{wmap5}. 
Recent Planck results are in agreement with this, with $A = 0.073 \pm 0.010$ in direction $(l, b) = (217.5, -20.2) \pm 15$ for the SMICA map and similar results for other maps \cite{planckasym}. However, on smaller scales the asymmetry is observed to be smaller. In particular, the asymmetry on scales corresponding to quasar number counts, $k \approx (1.3-1.8)h \;{\rm Mpc}^{-1}$, 
must satisfy $A < 0.012$ at 95$\%$ c.l. \cite{quasar}.

     An important question is whether this asymmetry could be due to a particular form of primordial perturbation, and whether there could be a plausible scalar field mechanism to generate this perturbation. There have been several recent proposals and analyses of this issue  \cite{pesky, recent1,isop1,recent2a,recent3,recent4,recent5,recent5a,recent6,recent7,recent8,recent9,dono}.

    Early proposals considered the possibility of spatial modulation of the CMB power in single-field inflation. However this is excluded by the associated CMB mean temperature quadrupole \cite{kam1}.  In \cite{kam1,kam2,kam3} it was proposed that the power asymmetry could be due to a mixture of isocurvature cold dark matter (CDM) and adiabatic perturbations from the decay of a curvaton. The curvaton is subdominant in order to suppress the hemispherical asymmetry in the energy density, which would otherwise lead to a large quadrupole in the CMB temperature. The scale-dependence of the isocurvature component may then allow the power asymmetry observed on large scales to be consistent with the lack of asymmetry in quasar number counts on smaller scales \cite{quasar}. In the case of WMAP5 constraints, it was shown that the model can account for the observed power asymmetry if the curvaton $\sigma$ has a large hemispherical power asymmetry, such that $\Delta \overline{\sigma}/\overline{\sigma} \sim 1$ across the present horizon \cite{kam3}\footnote{Here and in the following $\Delta \overline{\sigma}$ will denote the mean difference between the field $\sigma$ at the horizon and the mean value of the field over the volume corresponding to the observed Universe.}.
However, the tighter Planck constraints on the isocurvature perturbation and non-Gaussianity may exclude this model
 \cite{pesky}. In particular, the Planck constraint on 
non-Gaussianity strongly constrains the fraction of the adiabatic perturbation due to curvaton decay, making it more difficult to account for the asymmetry via a curvaton \footnote{In \cite{kam3} it was assumed that the curvaton perturbation was scale-independent. We have checked the case of a scale-dependent curvaton and find that it is only marginally possible to satisfy Planck bounds on non-Gaussianity and on the isocurvature perturbation while remaining consistent with the CMB quadrupole and quasar bound. This requires that all parameters are simultaneously close to their 2-$\sigma$ bounds. We will report on this analysis elsewhere \cite{jip}.}. 

   Since a strong constraint on the curvaton model is from non-Gaussianity, a scalar-field based explanation of the CMB asymmetry might best be achieved if non-Gaussianity can be suppressed. One way to achieve this is via a scale-dependent and spatially-modulated contribution to the total adiabatic perturbation from modulated reheating. As we will show, in the case where the dominant modulation of the inflaton decay rate is linear in the modulating field $\sigma$, it is possible to suppress $f_{NL}$ to below the Planck bound while accounting for the CMB power asymmetry.  

     For modulated reheating \cite{mr} to explain the CMB power asymmetry via a linear modulation, a specific form of modulating field perturbation is necessary. The perturbation must have an intrinsic hemispherical asymmetry and a red scale-dependence.   The red scale-dependence is essential to be able to account for the CMB power asymmetry on large scales while satisfying the quasar constraint on small scales. The modulating field perturbation must itself have an asymmetry, since the modulation of the inflaton decay rate is linear in the field. This is in contrast to the case of the curvaton, where the asymmetry in the CMB power is due to an asymmetry in the mean curvaton field rather than in the curvaton perturbation itself \cite{kam3}.

    As a specific example which can generate the required form of scale-dependent and asymmetric $\sigma$ perturbations, we will consider the  tachyonic growth model presented in \cite{isop1}. In this model, $\sigma$ is proportional to the phase of a complex field $\Sigma$ which undergoes tachyonic growth from an initial Bunch-Davies vacuum on sub-horizon scales at $\Sigma = 0$. Quantum fluctuations of $\sigma$ then acquire both a spatial modulation across our horizon and a red scale-dependence. We will show that all constraints can be satisfied in this model while accounting for the observed CMB power asymmetry.

   In Section 2 we review the tachyonic growth model for asymmetric and scale-dependent scalar field fluctuations. In Section 3 we discuss the CMB power asymmetry from modulated reheating in the tachyonic growth model and the observational constraints on the model. In Section 4 we present our results for the CMB power asymmetry and show that it is possible to satisfy all observational constraints. In Section 5 we discuss the modification of the spectral index and running spectral index due to
scale-dependent modulated reheating. In Section 6 we present our conclusions.

\section{The tachyonic growth model for modulating field fluctuations}

     We briefly review the tachyonic growth model of \cite{isop1}. This provides an explicit example which can generate the form of modulating field perturbations necessary to generate a CMB power asymmetry.

   The model is based on a complex scalar field $\Sigma \equiv (\Sigma_{o}/\sqrt{2}) e^{i \sigma/\Sigma_{o}}$. The potential is 
\be{n1} V(\Sigma) = -c H^{2} |\Sigma|^{2} + V_{lift}(\Sigma)  ~.\ee
Here $V_{lift}(\Sigma)$ contains the terms which determine the minimum of the potential. The $\Sigma$ field is assumed to be localized initially at $\Sigma = 0$, with a Bunch-Davies vacuum on subhorizon scales. $\Sigma$ then evolves in the tachyonic potential due to the first term in $V(\Sigma)$. In any horizon-sized volume, after a number of e-foldings $\Delta N$, there is a mean (rms) field, $\overline{\sigma}_{i}$ ($i = 1,\;2$), and a mean (rms) spatial variation of the field across the horizon, $\overline{\Delta \sigma}_{i}$, due to the net effect of superhorizon modes, where $\Sigma = (\sigma_{1} + i \sigma_{2})/\sqrt{2}$.  In \cite{isop1} it was found that $(\overline{\Delta \sigma}_{i}/\overline{\sigma}_{i})_{*} = 0.5$,  is obtained when $(\overline{\sigma}_{i}/H)_{*} \approx 1.5$ and $\Delta N$ is in the range 8.6 to 36.9 for $c$ in the range 1 to 0.1. Here $*$ denotes the time when our horizon exited the horizon during inflation. 

 The model can be considered to represent a generic phase transition occuring in a second field during inflation. $\Sigma$ will be localized at zero if its mass squared term is initially positive. It will then undergo a phase transition if its mass squared changes sign due to some model-dependent dynamics. Such transitions have been considered in the context of SUGRA models. For example, in \cite{adams}, phase transitions occur due to a negative mass squared term combined with a diminishing temperature correction. Alternatively, one could achieve such a phase transition by having two periods of inflation determined by two different inflaton fields. In this case the $O(H^2)$ mass squared term could change sign and become negative when the later period of inflation begins, since the sign of the $O(H^2)$ terms depends on the Kahler couplings of the two inflaton fields to $\Sigma$. In a non-SUSY context, such transitions could occur due to a non-minimal coupling of $\Sigma$ to gravity, $\xi R |\Sigma|^2$,  where $R = 12 H^{2}$. If $\xi$ were field dependent and changed sign, it could produce the necessary phase transition.

  In a given horizon volume, we can perform a field redefinition such that $\overline{\sigma}_{2} = 0$. $\sigma_{1}$ may then be considered the radial direction in the $\Sigma$ plane, while $\sigma_{2}$ is the angular direction for small $\sigma_{2}$, such that $\delta \theta = \delta \sigma_{2}/\overline{\sigma}_{1}$. The modulating field fluctuation is then $\delta \sigma = \Sigma_{o} \delta \theta$. $\delta \sigma_{2}$ at horizon exit is assumed to correspond to quantum fluctuations of a massless $\sigma_{2}$ field, with power spectrum $P_{\delta \sigma_{2}} = H^{2}/4\pi^{2}$. 
The scale-dependence of $\delta \sigma$ is due to the growth of $\overline{\sigma}_{1}$. This results in a red power spectrum for $\delta \sigma$ with spectral index $n_{\sigma} 
= 4 - \sqrt{4c + 9}$ \cite{isop1}. 
There is also a spatial modulation of the $\delta \sigma$ fluctuation due to spatial modulation of the radial field, $\overline{\Delta \sigma}_{1}$. This implies that $\delta \theta$ is also spatially modulated,
\be{n2} \delta \theta = 
\left( \frac{\delta \sigma_{2}}{\overline{\sigma}_{1}} \right)_{N} 
\frac{1}{
\left(1 + 
\frac{\Delta\overline{\sigma}_{1}}{\overline{\sigma}_{1}}
 \right)_{*} } ~,\ee
where $N$ is the number of e-foldings at which $\delta \sigma_{2}$ exits the horizon.
Therefore the modulation of all perturbations across our horizon is determined by the value of $\Delta \overline{\sigma}_{1}/\overline{\sigma}_{1}$ across our horizon when the perturbations exit the horizon. Since the ratio $\Delta \overline{\sigma}_{1}/\overline{\sigma}_{1}$ is constant for evolution in a $V \propto \sigma_{1}^2$ potential, its value across our horizon is fixed 
when our horizon exits the horizon during inflation. Hence
$(\overline{\Delta \sigma}_{1}/\overline{\sigma}_{1})_{*}$ determines the spatial modulation of all $\delta \sigma$ perturbations within our horizon. 

   In addition, there is a spatial modulation of the 
mean value of $\sigma$ across our horizon. This is due to the variation of 
$\sigma_{2}$ across our horizon, so that $\overline{\Delta 
\theta} \approx 
(\overline{\Delta \sigma_{2}}/\overline{\sigma}_{1})_{*}$ and 
$\overline{\Delta \sigma} = \Sigma_{o} \overline{\Delta 
\theta}$. As a result, the modulation of mean value of $\sigma$ across our horizon is of the same magnitude as the modulation of the $\delta \sigma$ power spectrum, since $\Delta \overline{\sigma}_{2} \approx \Delta \overline{\sigma}_{1}$.

    Thus $\sigma$ field perturbations from tachyonic growth have the two properties which are essential
for a modulated reheating explanation of the CMB power asymmetry: (i) the $\sigma$ field fluctuations are spatially modulated and (ii) the fluctuations have a red power spectrum. The first property is necessary because, as we will discuss, in order to have small non-Gaussianity, the modulation of the inflaton decay rate must be linear in the modulating field. Therefore spatial modulation of the energy density perturbations must come from direct spatial modulation of the modulating field perturbations. This is different from the case of the curvaton  \cite{kam3}, since in that case the curvaton energy density fluctuation is $\delta \rho \propto \sigma \times \delta \sigma$. Hence modulation of the mean value of $\sigma$  
across the horizon can modulate the power of the energy density fluctuations and so produce a CMB power asymmetry.  The second property is necessary to reduce the CMB power asymmetry at small angular scales and so satisfy the quasar constraint.

 \section{CMB hemispherical power asymmetry from modulated reheating}

\subsection{Inflaton decay in the tachyonic growth model}

      We will consider an interaction of the form 
\be{e18}         {\cal L}_{int}  \supset  - y(\Sigma) \phi \psi_{a} \psi_{a} \;\; + \;\; h.c.    ~.\ee 
Here $\phi$ is the inflaton and $\psi$ are fermions to which the inflaton decays. We expand the function 
$y(\Sigma)$ in a series 
\be{e19} y(\Sigma) =  y_{o}\left( 1 + \alpha \frac{\Sigma}{\Lambda} + \beta \frac{\Sigma^{2}}{\Lambda^{2}} + ... \right)  ~,\ee 
where $\Lambda$ is a mass scale large compared to $\Sigma$. 
Then the inflaton decay rate is 
\be{w1}  \Gamma = \Gamma_{o}\left(1 + \frac{\alpha \Sigma}{\Lambda}   +  \frac{\alpha^{*} \Sigma^{*}}{\Lambda} + ... \right)   ~.\ee  
With $\alpha = \alpha_{o}e^{i \delta} $ and $\Sigma = (\Sigma_{o}/\sqrt{2}) e^{i \sigma/\Sigma_{o}}$, the leading order decay rate becomes 
\be{w2}  \Gamma \approx \Gamma_{o} \left(1 + \frac{\sqrt{2} \alpha_{o} \Sigma_{o}}{\Lambda}\left(\cos\left(\frac{\sigma}{\Sigma_{o}} + \delta\right) \right) \right) ~.\ee
Without loss of generality we can choose the minimum of $V(\sigma)$ to be at $\sigma = 0$ by redefining the phase $\delta$. 

    As we will discuss later, in order to suppress non-Gaussianity it is necessary for $\sigma$ to be damped down from its initial value,  such that $\sigma/\Sigma_{o}$ is small compared with 1. In this case we can expand the decay rate as 
\be{w3}  \Gamma \approx \Gamma_{o}\left(1 + \frac{\tilde{\alpha} \sigma}{\Sigma_{o}}  + \frac{\tilde{\beta} \sigma^{2}}{\Sigma_{o}^{2}} + ... \right) ~,\ee
where
\be{w4} \tilde{\alpha} = -\sqrt{2} \alpha_{o} \frac{\Sigma_{o}}{\Lambda} \sin(\delta) \;\;\; ; \;\;\;  \tilde{\beta} = -\sqrt{2} \alpha_{o} \frac{\Sigma_{o}}{2 \Lambda} \cos(\delta) ~.\ee

    To achieve the damping of $\sigma$, we need to break the global $U(1)$ symmetry of $V(\Sigma)$ and generate a potential for $\sigma$. As a simple example, suppose that
\be{n3}  V_{lift}(\Sigma) = \frac{\lambda}{4} |\Sigma|^{4} + 
\frac{\gamma}{4} \left(\Sigma^{4} + \Sigma^{*\;4}\right)   ~,\ee
where $\gamma \ll \lambda$ and for simplicity we consider $\gamma$ to be real. The minimum of the $\Sigma$ potential in the limit $\gamma \ll \lambda$ is at
\be{n4} |\Sigma|^{2} \equiv \frac{\Sigma_{o}^{2}}{2} \approx\frac{2cH^{2}}{\lambda}    ~.\ee 
The potential for $\sigma$ is then
\be{n5}  V(\sigma) \approx \frac{\gamma \Sigma_{o}^{4}}{8} \cos \left( \frac{4 \sigma}{\Sigma_{o}} \right)  ~.\ee 
For $4 \sigma$ small compared with $\Sigma_{o}$ this becomes 
\be{n6} V(\sigma) \approx -\gamma \Sigma_{o}^{2} \sigma^{2} + \;\; {\rm constant}  ~.\ee  
$\gamma < 0$ is necessary since we have defined the minimum of the potential to be at $\sigma = 0$. Then 
\be{n8} V(\sigma) \approx \frac{4 c |\gamma|}{\lambda} H^{2} \sigma^{2}    ~.\ee
Thus $\sigma$ will undergo damped evolution towards $\sigma = 0$ if $4 c |\gamma|/\lambda < 1$. This results in damping of $\sigma$ by a factor $f_{d}$ from its initial value $\sigma_{o}$. It is important that any evolution of $\sigma$ is damped and not rapidly oscillating, since if it were oscillating then the linear term in $\Gamma(\sigma)$ would average to zero and so could not modulate the inflaton decay rate.

\subsection{Adiabatic perturbation from modulated reheating}

   The contribution to the total adiabatic curvature perturbation $\zeta$ from modulated reheating is \cite{ichikawa}
\be{w10}  \zeta_{MR} =  Q_{\sigma}\delta \sigma + \frac{1}{2} Q_{\sigma \sigma} \delta \sigma^{2} + ...    ~,\ee
where 
\be{w10a} Q_{\sigma} = A \Gamma_{\sigma}/\Gamma \;\;;\;\; Q_{\sigma \sigma} = A \Gamma_{\sigma \sigma}/\Gamma + B \left(\Gamma_{\sigma}/\Gamma\right)^{2}  ~.\ee
 Here $\delta \sigma$ is the fluctuation of $\sigma$ at the time of inflaton decay \footnote{In \cite{ichikawa}, $\delta \sigma$ is the value at horizon exit. However, in the model of \cite{ichikawa} there is no evolution of $\delta \sigma$ for superhorizon perturbations, since $\sigma$ is assumed to be massless. Therefore $\delta \sigma$ is also the perturbation at inflaton decay. In our model $\delta \sigma$ evolves between horizon exit and inflaton decay. In this case $\delta \sigma$ should be interpreted as the value at the time of inflaton decay.}.  The inflaton is assumed to have a potential of the form $V(\phi) \propto \phi^{2n}$ near the minimum. $A$ and $B$ are constants which depend on the value of $n$. For $n = 1$, as we conventionally expect, $A = -1/6$  and $B = 1/6$ \cite{ichikawa}. 

   Thus with $A = 1/6$ and $B = -1/6$, and using the expansion \eq{w3}, we obtain 
\be{n20} \zeta_{MR} \approx -\frac{1}{6} \frac{\Gamma_{\sigma}}{\Gamma} \delta \sigma \approx  -\frac{1}{6} \frac{\tilde{\alpha}}{\Sigma_{o}} \delta \sigma  ~.\ee
The power spectrum of $\zeta_{MR}$ is then 
\be{n22}  P_{\zeta_{MR}} = \frac{1}{36} 
\frac{\tilde{\alpha}^{2}}{\Sigma_{o}^{2}} P_{\delta \sigma}  ~.\ee
$\delta \sigma$ is related to $\delta \theta$ by $\delta \sigma = \Sigma_{o} \delta \theta$, where 
\be{n23}  \delta \theta = \left(\frac{\delta \sigma_{2}}{\overline{\sigma}_{1}} \right)_{N} \frac{ f_{d} }{ \left( 1 + \frac{\Delta \overline{\sigma}_{1}}{\overline{\sigma}_{1}}\right)_{*} }   ~.\ee
In this we have included a damping factor $f_{d}$ for the $\delta \theta$ perturbation relative to its initial value from the tachyonic growth model. Therefore 
\be{n24}  P_{\delta \sigma} = \left(\frac{P_{\delta \sigma_{2}}}{\overline{\sigma}_{1}^{2}} \right)_{N} \frac{f_{d}^{2}   
 \Sigma_{o}^{2}}{ 
\left( 1 + 
\frac{\Delta \overline{\sigma}_{1}}{\overline{\sigma}_{1}}\right)_{*}^{2} 
}   ~.\ee
Defining $N_{o}$ to be the number of e-foldings corresponding to the pivot scale $k_{o}$ of $P_{\zeta_{MR}}$, using $P_{\delta \sigma_{2}} = H^{2}/4 \pi^{2}$ and setting $\Delta \overline{\sigma_{1}}$ to zero for now, we obtain 
\be{n25} P_{\delta \sigma} = \frac{\Sigma_{o}^{2} f_{d}^{2}}{4 \pi^{2}} 
\left( \frac{H}{\overline{\sigma}_{1}} \right)^{2}_{N_{o}} 
\left( \frac{k}{k_{o}} \right)^{n_{\sigma} - 1}  
~\ee
and so
\be{n25a} P_{\zeta_{MR}} = \frac{\tilde{\alpha}^{2} f_{d}^{2}}{144 \pi^{2}} 
\left( \frac{H}{\overline{\sigma}_{1}} \right)^{2}_{N_{o}} 
\left( \frac{k}{k_{o}} \right)^{n_{\sigma} - 1}  
~.\ee
Therefore, defining $\xi_{o} \equiv P_{\zeta_{MR}}(k_{o})/P_{\zeta}$, and using $P_{\zeta} \approx P_{\zeta\; inf}$, where $P_{\zeta\;inf}$ is the power spectrum of the dominant inflaton adiabatic perturbation, 
we obtain
\be{n26}  \xi_{o} \approx \frac{P_{\zeta_{MR}}(k_{o})}{P_{\zeta\;inf}} = 
\frac{\tilde{\alpha}^{2}}{36 {\cal A}^{2}} \frac{f_{d}^{2}}{4 \pi^{2}} 
\left( \frac{H}{\overline{\sigma}_{1}} \right)^{2}_{N_{o}}   ~,\ee
where ${\cal A} = 4.8 \times 10^{-5}$ is the amplitude of the observed adiabatic curvature perturbation.

The value of 
$\tilde{\alpha}$ necessary to account for a given value of $\xi_{o}$ is therefore 
\be{n27} \tilde{\alpha} \approx \frac{12 \pi \xi_{o}^{1/2} {\cal A}}{ 
f_{d} \left( \frac{H}{\overline{\sigma}_{1}} \right)_{N_{o}}  
}  ~.\ee
Thus, to generate a given $\xi_{o}$,  $\tilde{\alpha}$ must satisfy 
\be{n28} \tilde{\alpha} \approx \frac{5.7 \times 10^{-4}}{f_{d}} 
\left( \frac{\xi_{o}}{0.1} \right)^{1/2} 
\left( \frac{\overline{\sigma}_{1}}{H} \right)_{N_{o}} 
 ~.\ee

\subsection{Non-Gaussianity constraint}

     For the case of decay to a pair of fermions, the local non-Gaussianity in modulated reheating is \cite{ichikawa}
\be{e21}  \frac{6}{5} f_{NL} = \frac{R^{2}}{A\left(1 + R\right)^{2}} \left( \frac{B}{A} +  
\frac{\Gamma \Gamma_{\sigma \sigma}}{\Gamma_{\sigma}^{2}} \right)   ~.\ee
$R$ in \eq{e21} is defined to be the ratio of the square of the curvature perturbation from modulated reheating to that from the inflaton, $\zeta_{MR}^{2}/\zeta_{inf}^{2}$. Therefore $R \approx \xi$ in the absence of scale-dependence of $\xi$, since $\xi = P_{\zeta_{MR}}/P_{\zeta} \approx P_{\zeta_{MR}}/P_{\zeta\;inf}$ in that case. When computing $f_{NL}$ we will set $\xi$ equal to $\xi_{o}$, where the pivot scale is $k_{o} = 0.002 {\rm Mpc}^{-1}$.  

Thus, assuming that $\xi \ll 1$, as necessary for a subdominant modulated reheating contribution, we have  
\be{e22} f_{NL}  \approx \frac{-5 \xi^{2}}{\left(1 + \xi\right)^{2}} \left(-1 + \frac{\Gamma \Gamma_{\sigma \sigma}}{\Gamma_{\sigma}^{2}} \right)   ~.\ee
Using \eq{w3} for $\Gamma(\sigma)$, we obtain
\be{e23} \frac{\Gamma \Gamma_{\sigma \sigma}}{\Gamma_{\sigma}^{2}} = \frac{ 2 \tilde{\beta}}{\left(\tilde{
\alpha} + \frac{2 \tilde{\beta} \sigma}{\Sigma_{o}}   \right)^{2}} ~.\ee
We see here why the linear term in the inflaton decay rate should dominate in order to minimize non-Gaussianity. If $\tilde{\alpha} \rightarrow 0$ in \eq{e23}, then this term is $O(\Sigma_{o}^{2}/\sigma^{2})/\tilde{\beta} \gg 1/\tilde{\alpha}$ (using $|\tilde{\alpha}| \sim |\tilde{\beta}|$) rather than $O(\tilde{\beta}/\tilde{\alpha}^{2}) \sim 1/\tilde{\alpha}$. Hence $f_{NL}$ would be greatly enhanced.

With $|\sigma/\Sigma_{o}| \ll 1$, and using $\xi \ll 1$ and $|\tilde{\alpha}| \sim |\tilde{\beta}|$ if $|\tan \delta| \sim 1$, we find 
\be{n9} f_{NL} \approx 5 \xi^{2} \left(1 - \frac{2 \tilde{\beta}}{\tilde{\alpha}^{2}} \right)   ~.\ee

Planck imposes the constraint $f_{NL} = 2.7 \pm 5.8$ (1-$\sigma$) \cite{planckNG}. Thus the 2-$\sigma$ upper bound on $|f_{NL}|$ is $|f_{NL\;lim}| = 14.3$.
We can assume that $5 \xi^2 \ll 1$, since we will show that $\xi_{o} \sim 0.1$ is necessary for the model to satisfy all constraints. Therefore the first term in \eq{n9} will satisfy the Planck bound on $|f_{NL}|$. Thus 
\be{n10} f_{NL} \approx - \frac{5 \xi^{2}_{o}}{\tan(\delta) \tilde{\alpha}}    ~,\ee
where we have used $\tilde{\beta} = \tilde{\alpha}/2 \tan(\delta)$. Therefore $|f_{NL}| < |f_{NL\;lim}|$ requires that
\be{n11} \tilde{\alpha} \gae \frac{5 \xi^{2}_{o}}{|\tan(\delta)| |f_{NL \; lim}|}   ~.\ee
Thus the non-Gaussianity constraint on $\tilde{\alpha}$ is 
\be{n11a} \tilde{\alpha} \gae 3.6 \times 10^{-3} \frac{1}{|\tan(\delta)|} \left(\frac{14}{|f_{NL\;lim}|}\right) \left(\frac{\xi_{o}}{0.1}\right)^{2} ~.\ee

\subsection{CMB Quadrupole constraint}

   We will apply the method of \cite{kam2} to the case of modulated reheating. In \cite{kam2}, the superhorizon curvaton fluctuation responsible for spatial modulation of the CMB power is assumed to have a sinusoidal form with a single wavenumber $\bfk$,
\be{w6}  \sigma = \overline{\sigma}  + \overline{\sigma}_{k} \sin\left(\bfk.\bfx   + \omega_{o} \right)    ~.\ee

   In the tachyonic growth model, the spatial modulation of $\sigma$ is due to a sum of superhorizon modes of $\sigma_{2}$. The sum is dominated by modes which are not much larger than the horizon when our universe exits the horizon during inflation. This can be seen since the integral for $\overline{\Delta \phi}$ in \cite{isop1} (Eq. (3.29) of \cite{isop1}) is dominated by modes close to the upper bound, $k_{max}$, corresponding to $\lambda_{phys} \sim  (0.1-1) H^{-1}$. 

   In order to apply the method of \cite{kam2}, we will model the superhorizon fluctuation of the modulating field by a single mode with wavelength close to the horizon at horizon exit, $kx_{dec} \sim 0.1$, where $x_{dec}$ is the comoving distance to the last-scattering surface. 

   It is assumed in \cite{kam2} that the superhorizon perturbation of $\sigma$ results in a gravitational (Bardeen)
 potential perturbation $\Psi$ which can be expanded in the form 
\be{w8}  \Psi = \Psi_{k}(\tau_{dec}) \left[ \sin \omega_{o} + \cos \omega_{1} \left(\bfk.\bfx\right) - \frac{\sin \omega_{2}}{2} \left(\bfk.\bfx\right)^{2}  -  \frac{\cos \omega_{3}}{6} \left(\bfk.\bfx\right)^{3}  + O((\bfk.\bfx)^{4}) \right]   ~,\ee
where $\tau$ is conformal time and $|\bfk.\bfx| \ll 1$. 
The leading order contribution to the quadrupole $a_{20}$ is then \cite{kam2} 
\be{w9} a_{20} = -\sqrt{\frac{4 \pi}{5}} \left( k x_{dec}\right)^{2} \delta_{2} \frac{\sin \omega_{2}}{3} \Psi_{\bfk}(\tau_{dec})    ~,\ee
where $\delta_{2} = 0.33$ for a $\Lambda$CDM universe. $\Psi_{\bfk}(\tau_{dec})$ is related to the primordial gravitational perturbation (the perturbation during radiation domination), $\Psi_{\bfk}$, by $\Psi_{\bfk}(\tau_{dec}) = 0.937 \Psi_{\bfk}$.  
$a_{20}$ should be less than the observational upper bound, $Q$. In \cite{kam2} the upper bound is assumed to be three times the variance of the quadrupole, therefore  $Q = 3 \sqrt{C_{2}} = 1.8 \times 10^{-5}$.

     To compute the quadrupole for the modulated reheating model, we therefore need the gravitational potential $\Psi$
for a sinusoidal superhorizon modulating field
fluctuation  of the form \eq{w6}. The primordial gravitational potential perturbation due to modulated reheating is given by $\Psi = -2 \zeta_{MR}/3$, where $\zeta_{MR}$ is given by \eq{n20}, therefore 
\be{w11} \Psi \approx  \frac{1}{9} \frac{\Gamma_{\sigma}}{\Gamma} \delta \sigma     ~.\ee
Thus 
\be{w13} \Psi \approx  \frac{1}{9} \frac{\tilde{\alpha}}{\Sigma_{o}} 
\overline{\sigma}_{\bfk} \sin\left(\bfk.\bfx  + \omega_{o}\right)   ~.\ee 
 Expanding this in $\bfk.\bfx$ gives the term responsible for the quadrupole,  
\be{w14} \Psi = - \frac{1}{18} \frac{\tilde{\alpha} \overline{\sigma}_{\bfk}}{\Sigma_{o}} (\bfk.\bfx)^{2} \sin \omega_{o} + ...    ~.\ee   
Comparing with \eq{w8}, we find 
\be{w15}  \Psi_{\bfk} \sin \omega_{2} \equiv   \frac{1}{9} \frac{\tilde{\alpha} \overline{\sigma}_{\bfk}}{\Sigma_{o}}  \sin \omega_{o} ~.\ee 
The quadrupole upper bound requires that \cite{kam2} 
\be{w16} 0.937 (k x_{dec})^{2} |\Psi_{k} \sin \omega_{2}| \lae 5.8 Q   ~,\ee
where we have included the correction factor relating $\Psi_{\bfk}$ to $\Psi_{\bfk}(\tau_{dec})$.  
Therefore, with $\Delta \overline{\sigma} = (k x_{dec}) \overline{\sigma}_{k}$, we obtain  
\be{w17} \frac{\Delta \overline{\sigma}}{\Sigma_{o}} \frac{\tilde{\alpha}}{9} (kx_{dec}) \sin \omega_{o} \lae 6.2 Q   ~.\ee
In general $\theta = \overline{\sigma}/\Sigma_{o}$, therefore 
\be{w17a} \frac{\Delta \overline{\sigma}}{\overline{\sigma}} \frac{\tilde{\alpha}}{9} (kx_{dec}) \theta \sin \omega_{o} \lae 6.2 Q   ~.\ee
Initially $\overline{\sigma} = \overline{\sigma}_{o} = \Sigma_{o} \theta_{o}$, where $\theta_{o}$ is random and so typically $\theta_{o} \sim 1$. Subsequently $\overline{\sigma}$ is damped
as the $\sigma$ field  evolves towards the minimum of $V(\sigma)$ at $\sigma = 0$. Therefore $\overline{\sigma} =  f_{d} \overline{\sigma}_{o}$ and $\theta = f_{d} \theta_{o} \sim f_{d}$. In addition, $\Delta \overline{\sigma}/\overline{\sigma} \approx \Delta \overline{\sigma}_{1\;*}/\overline{\sigma}_{1\;*}$.  Therefore the condition to satisfy the quadrupole bound ($Q = 1.8 \times 10^{-5}$) is
\be{w18} \tilde{\alpha} \lae \frac{1.0 \times 10^{-2}}{\left( \frac{\Delta \overline{\sigma}_{1}}{\overline{\sigma}_{1}}\right)_{*} (\frac{k x_{dec}}{0.1}) \sin \omega_{o} f_{d} }   ~.\ee

\subsection{Calculation of the hemispherical CMB power asymmetry}

    The spatial modulation of the CMB power asymmetry is due to the spatial modulation of the power spectrum of $\zeta_{MR}$, 
\be{s1} \frac{\Delta C_{l}}{C_{l}} =  \frac{\Delta P_{\zeta_{MR}}}{P_{\zeta}} = \frac{P_{\zeta_{MR}}}{P_{\zeta}} \frac{\Delta P_{\zeta_{MR}}}{P_{\zeta_{MR}}} \approx \xi  \frac{\Delta P_{\zeta_{MR}}}{P_{\zeta_{MR}}}   ~\ee
Here $\xi$ is $l$-dependent due to scale-dependence of the 
adiabatic perturbation from modulated reheating. Since 
\be{s2} P_{\zeta_{MR}} \propto P_{\delta \sigma} 
\propto \frac{1}{
\left(1 + \frac{\Delta \overline{\sigma}_{1}}{\overline{\sigma}_{1}} \right)_{*}^{2} 
}  ~.\ee
we obtain
\be{s3}  
\frac{\Delta C_{l}}{C_{l}} = \xi \times \frac{ \Delta P_{\zeta_{MR}} }{P_{\zeta_{MR}} } = 
\xi \times \frac{1 - \left(1 + \frac{\Delta \overline{\sigma}_{1}}{\overline{\sigma}_{1}} \right)_{*}^{2}  
}{
\left(1 + \frac{\Delta \overline{\sigma}_{1}}{\overline{\sigma}_{1}} \right)_{*}^{2} 
}
~,\ee
where $\Delta P_{\zeta_{MR}} = P_{\zeta_{MR}}(\Delta \overline{\sigma}_{1\;*}) -  P_{\zeta_{MR}}(\Delta \overline{\sigma}_{1\;*} = 0)$. Therefore 
\be{s4}  \left|\frac{\Delta C_{l}}{C_{l}}\right| = 2 
\left|\frac{\Delta \overline{\sigma}_{1\;*}}{\overline{\sigma}_{1\;*}}\right| \kappa \xi
~,\ee    
where 
\be{s4x}   \kappa = \left| \frac{1 - \left(1 + \frac{\Delta \overline{\sigma}_{1}}{\overline{\sigma}_{1}} \right)_{*}^{2}  
}{2 
\left|\frac{\Delta \overline{\sigma}_{1\;*}}{\overline{\sigma}_{1\;*}}\right|
\left(1 + \frac{\Delta \overline{\sigma}_{1}}{\overline{\sigma}_{1}} \right)_{*}^{2} } \right|   ~.\ee

For $|\Delta \overline{\sigma}_{1\;*}/\overline{\sigma}_{1\;*}| \ll 1$, $\kappa \rightarrow 1$. The value of $\Delta \overline{\sigma}_{1\;*}/\overline{\sigma}_{1\;*}$ from the tachyonic growth model is an rms magnitude. In a given horizon volume, $\Delta \overline{\sigma}_{1\;*}/\overline{\sigma}_{1\;*}$ can enter \eq{s3} within a range of values of positive or negative sign. For example, we will 
be interested in the case $\Delta \overline{\sigma}_{1\;*}/\overline{\sigma}_{1\;*} = 0.5$ from the tachyonic growth model. 
If this enters \eq{s3} with a positive sign, then 
$\Delta C_{l}/C_{l} = -5 \xi/9$, whereas if it enters with a negative sign then $\Delta C_{l}/C_{l} = 3 \xi$. We will therefore use negative $\Delta \overline{\sigma}_{1\;*}/\overline{\sigma}_{1\;*}$ in order to maximize the asymmetry.

   The hemispherical power asymmetry on large scales ($l \; \leq \; l_{max} = 64$) is obtained in \cite{kam3} by averaging over the individual modes. The asymmetry $A$ on large scales, which we denote by $A_{large}$, is then given by 
\be{e1} A_{large} \equiv \frac{\Delta \overline{\sigma}_{1\;*}}{\overline{\sigma}_{1\;*}} \tilde{A} ~,\ee
where $\tilde{A}$ is defined by
\be{e1a} \tilde{A} = \sum_{l=2}^{l_{max}} \frac{2l+1}{\left(l_{max} - 1\right)\left(l_{max} + 3\right)} K_{l} ~,\ee
and $K_{l}$ is defined by 
\be{e2}   \left| \frac{\Delta C_{l}}{C_{l}} \right| \equiv 2 \frac{\Delta \overline{\sigma}_{1\;*}}{\overline{\sigma}_{1\;*}} K_{l}   ~.\ee
Therefore $K_{l} = \kappa \xi$ in the modulated reheating model. 
The $l$-dependence of $K_{l}$ is due to the scale-dependence of $\xi$, 
\be{s4a} \xi = \left(\frac{k}{k_{o}}\right)^{n_{\sigma} - 1} \xi_{o}   ~,\ee
where the spectral index is $n_{\sigma} = 4 - \sqrt{4c + 9}$ in the tachyonic growth model. We will use $c$ to parameterize the scale-dependence in the following. For a given $k$ the corresponding multipole is  $l \approx k d_{A}^{c}$, where $d_{A}^{c}$ is the comoving angular distance scale, $d_{A}^{c} \approx 2 H_{0}^{-1}/\Omega_{m}^{0.4}$. With $H_{o}^{-1} \approx 3000 h^{-1} \; {\rm Mpc}$, $h = 0.68$ and $\Omega_{m} = 0.30$ we obtain $l \approx 14100 \;k \; {\rm Mpc}$. The pivot scale, which we 
define to be $k_{o} = 0.002 \; {\rm Mpc}^{-1}$, then corresponds to $l_{o} \approx 28$. We can then approximate the scale-dependence factor by 
\be{e9}   \left( \frac{k}{k_{o}} \right)^{n_{\sigma} - 1} \approx  \; \left( \frac{l}{l_{o}} \right)^{n_{\sigma} - 1}   ~.\ee
The range of $l$ relevant to quasar number counts, 
$k = (1.3-1.8) h {\rm Mpc^{-1}} $, corresponds to $l \approx 12400-17200$. We can set $l = 15000$ for all $l$ when averaging over $K_{l}$ in this range, as $K_{l}$ does not vary much over this range of $l$. Then $\tilde{A}$ on quasar number count scales is 
\be{s5} \tilde{A} \approx \kappa \xi(l = 15000)      ~.\ee
The small-scale asymmetry on quasar scales is then 
\be{s6} A_{small} \approx \frac{\Delta \overline{\sigma}_{1\;*}}{\overline{\sigma}_{1\;*}}  \kappa \xi(l = 15000)    ~.\ee

\section{Results}

   In Table 1 we show the values of $\xi_{o}$ as a function of $c$ for which the large scale asymmetry is equal to the observed mean value, $A_{large}= 0.072$, for the case $\Delta \overline{\sigma}_{1\;*}/\overline{\sigma}_{1\;*} = 0.5$. We also show the corresponding value of $A_{small}$. We find that $c \geq 0.5$ is necessary in order to have a strong enough scale-dependence to  
satisfy the quasar bound $A_{small} < 0.012$. This corresponds to a red spectral index for the modulating field perturbations, $n_{\sigma} < 0.683$. The values of $\xi_{o}$ are in the range 0.049 to 0.062 for $c$ in the range 0.5 to 1.0. Thus we typically require $\xi_{o} \approx 0.05 - 0.1$ to be able to account for the CMB power asymmetry.

   We next consider whether it is possible to generate $\xi_{o} \sim 0.1$ in the tachyonic growth model while remaining consistent with the constraints from non-Gaussianity and the CMB quadrupole.

  We first consider the constraint from non-Gaussianity. With $\xi_{o} \sim 0.1$, we find that the non-Gaussianity lower bound on $\tilde{\alpha}$, \eq{n11a}, can be safely satisfied if $\tilde{\alpha} \gae 10^{-2}$.

We next consider the constraint from requiring that $\xi_{o} \sim 0.1$ can be generated via tachyonic growth. In the tachyonic growth model, $\Delta \overline{\sigma}_{1\;*}/\overline{\sigma}_{1\;*} = 0.5$ is achieved when $(\overline{\sigma}_{1}/H)_{*} \approx 1.5$ \cite{isop1}. 
The requirement that $\xi_{o} \sim 0.1$ can be generated, \eq{n28}, then requires that 
$\tilde{\alpha} \approx 9 \times 10^{-4}/f_{d}$. In order to be safely consistent with the bound from non-Gaussianity, we therefore require some damping of the modulating field, $f_{d} \lae 0.1$. 

    Finally, we consider the CMB quadrupole constraint. The quadrupole bound \eq{w18}  requires that $\tilde{\alpha} \lae 10^{-2}/f_{d}$. Thus damping by $f_{d} \lae 0.1$ also allows the quadrupole to be easily consistent with non-Gaussianity. 

   Therefore, in the context of the tachyonic growth model for modulating field perturbations, it is possible to generate the observed hemispherical CMB power asymmetry via modulated reheating while remaining consistent with all observational constraints. 

\begin{table}[h]
\begin{center}
\begin{tabular}{|c|c|c|}
 \hline $c$	 &	$\xi_{o}$  & $A_{small}$  \\
\hline	$0.0$	&	$0.049$	&	$0.072$   \\
\hline	$0.2$	&	$0.052$	&	$0.034$    \\
\hline	$0.4$	&	$0.055$	&	$0.016$   	\\
\hline	$0.49$	&	$0.0558$	&	$0.012$   	\\
\hline	$0.6$	&	$0.057$	&	$0.0081$    \\
\hline	$0.8$	&	$0.060$	&	$0.0040$    \\
\hline	$1.0$	&	$0.062$	&	$0.0021$    \\
\hline     
 \end{tabular} 
 \caption{\footnotesize{$\xi_{o}$ and $A_{small}$ as a function of $c$ when $A_{large} = 0.072$ and $\Delta \overline{\sigma}_{1\;*}/\overline{\sigma}_{1\;*} = 0.5$.}  }
 \end{center}
 \end{table}

\section{Modified spectral index and its running due to scale-dependent modulated reheating}

 The introduction of a scale-dependent modulated reheating component of the adiabatic density perturbation will modify the CMB spectral index as compared with the pure inflaton perturbation. This modification is itself scale-dependent, decreasing on smaller scales, therefore a running spectral index is predicted.  We can write the total curvature perturbation power spectrum as  
\be{e18}    P_{\zeta} = P_{\zeta_{inf}} + P_{\zeta_{MR}} = P_{\zeta_{inf}}\left(1 + \xi_{o} \left( \frac{k}{k_{o}}\right)^{n_{\sigma} - 1}    \right)  ~.\ee 
The spectral index $n_{s}$ of the total adiabatic perturbation is 
\be{e19} n_{s} - 1 = \frac{1}{P_{\zeta}} \frac{d P_{\zeta}}{d \ln k}   ~.\ee
Therefore the shift in the spectral index due to the modulated reheating component is 
\be{e20}  \Delta n_{s} \approx \xi_{o} \left( n_{\sigma} - 1 \right) \left(\frac{k}{k_{o}}\right)^{n_{\sigma} - 1}   ~,\ee
where we assume $\xi_{o} \ll 1$. The running spectral index due to modulated reheating is then 
\be{e20a}  n^{\prime} \equiv \frac{ d n_{s}}{d \ln k} = \xi_{o} \left( n_{\sigma} - 1 \right)^{2} \left(\frac{k}{k_{o}}\right)^{n_{\sigma} - 1} ~.\ee

   For example, the limiting case from Table 1 with $A_{large} = 0.072$ and $A_{small} = 0.012$ corresponds to $\xi_{o} = 0.0558$ and $c = 0.49$. The corresponding modulating field spectral index is $n_{\sigma} = 0.689$. The shift in the spectral index at $k_{o}$ is then $\Delta n_{s} = -0.0174$, while the running spectral index is
\be{e21a}  n^{\prime} = 0.005 \left(\frac{k}{k_{o}}\right)^{-0.31}   ~.\ee
The Planck 68 $\%$ c.l. bound on the running spectral index at $k = 0.05 {\rm Mpc}^{-1}$ is  $n^{\prime} =  -0.013 \pm 0.009$ 
\cite{planckcosmo}. Thus the 2-$\sigma$ upper bound on $n^{'}$ at $k = 0.05 {\rm Mpc}^{-1}$ is 0.005. With $k_{o} = 0.002 {\rm Mpc}^{-1}$, we find from \eq{e21} that the running spectral index due to the modulated reheating component at $k = 0.05 {\rm Mpc}^{-1}$ is $n^{\prime} = 0.002$. Thus, in the case where the running of the spectral index due to the inflaton component of the adiabatic perturbation is negligible, the total running spectral index is within the Planck 2-$\sigma$ range. Therefore observation of a small positive running spectral index would be consistent with the modulated reheating and tachyonic growth model for the CMB power asymmetry.

  The modulated reheating component can also significantly modify the predictions of common inflation models. As an explicit example, consider the case of hybrid inflation with a logarithmic potential, as in SUSY hybrid inflation. In this case the inflaton produces a spectral index $n_{s\;inf} = 1 - 1/N = 0.983$ ($N = 60$). The running spectral index is $n^{\prime} = -1/N^{2} \sim - 3 \times 10^{-4}$ and is therefore negligible. 
For the case $c = 0.5$ and $\Delta \overline{\sigma}_{1\;*}/\overline{\sigma}_{1\;*} = 0.5$, the total spectral index at $k_{o}$ is then $n_{s\; inf} + \Delta n_{s} = 0.966$, with a positive running spectral index $n^{\prime} = 0.002$ at $k = 0.05 {\rm Mpc^{-1}}$. We note that the total spectral index is in good agreement with the Planck value, $n_{s} = 0.9603 \pm 0.0073$ \cite{planckcosmo}, in contrast to the spectral index of the original hybrid inflation model.

\section{Conclusions}

   We have shown that the hemispherical asymmetry of the CMB temperature fluctuations can be explained via a subdominant modulated reheating contribution to the adiabatic perturbation. In general, the modulating field must have a power spectrum which has an intrinsic hemispherical asymmetry and is scale-dependent with a red spectrum.  The inflaton decay rate must be dominated by a term linear in the modulating field, in order to suppress non-Gaussianity. The red spectrum is then essential to suppress the asymmetry at small scales and so evade the 
constraint from quasar number counts.  

     The form of modulating field perturbation necessary to account for the hemispherical asymmetry can be generated via tachyonic growth of a complex scalar field from an initial Bunch-Davies vacuum on subhorizon scales at $\Sigma = 0$. This can produce both the required red spectrum for the CMB power asymmetry and a large hemispherical asymmetry in the modulating field, where the modulating field is proportional to the phase of $\Sigma$ in this model. We find that it is possible to account for the CMB power asymmetry while satisfying the quasar bound, the Planck upper bound on $f_{NL}$ and the upper bound on the CMB quadrupole. This requires that modulated reheating contributes approximately 5-10$\%$ of the total adiabatic power at large angular scales.

The tachyonic growth model requires some specific features and conditions to be satisfied. The complex scalar field must be initially localized at $\Sigma = 0$. Our horizon must then exit the horizon during inflation while the $\Sigma$ potential is dominated by the tachyonic term. Finally, the modulating field must have a potential such that it undergoes damped evolution towards its minimum, in order to satisfy the non-Gaussianity constraint. 

       The condition that the field is still undergoing tachyonic evolution requires that the phase transition initiating tachyonic growth occurs only 10-40 e-foldings before our horizon exits the horizon during inflation. This  is a typical requirement of any physical process which can generate superhorizon perturbations with observable effects on the scale of our horizon, in order that such superhorizon effects are not either stretched or damped by inflation to become unobservable.  
      
      A second possible issue is the effect of the modulating field $\Sigma$ on the inflaton dynamics. It is possible that the inflaton and $\Sigma$ could behave as a two-field inflation model. This will depend on the model-dependent couplings of the inflation field and $\Sigma$.

   A prediction of this class of model is that there will be a shift of the spectral index and a small positive running spectral index due to the modulated reheating component of the adiabatic perturbation. For the tachyonic growth model with $c = 0.5$, we find $\Delta n_{s} = -0.0174$ and $n^{\prime} = 0.002$ at $k = 0.05 {\rm Mpc}^{-1}$. The running spectral index due to modulated reheating is within the Planck 2-$\sigma$ upper bound $n^{\prime} = 0.005$. Thus if inflaton contribution to the running spectral index is negligible, then observation of a small positive running spectral index would be consistent with the tachyonic growth model. More generally, we expect a modification of the predictions of common inflation models. For example, the predictions for a logarithmic potential hybrid inflation model are $n_{s} = 0.983$ and $n^{\prime} = - 3 \times 10^{-4}$. In the tachyonic growth model for the asymmetry with $c = 0.5$, these become $n_{s} = 0.966$ and $n^{'} = 0.002$.

    The model provides an example where a scalar field can generate the CMB power asymmetry. It may therefore provide some insight into the general conditions necessary to achieve this via a scalar field. In particular, even though the model seeks to minimize non-Gaussianity, non-Gaussianity nevertheless imposes a significant constraint, requiring some field dynamics to bring it into line with the generation of a sufficiently large modulated reheating perturbation and the CMB quadrupole. This illustrates the importance of non-Gaussianity as a constraint on scalar field models of the CMB power asymmetry.

 \section*{Acknowledgements}
The work of JM is supported by the Lancaster-Manchester-Sheffield Consortium for Fundamental Physics under STFC grant
ST/J000418/1.

\end{document}